\documentclass[prb, reprint, showpacs, superscriptaddress]{revtex4-1}

\usepackage[version=3]{mhchem}
\usepackage{mathtools}
\usepackage{gensymb}
\usepackage{hhline}
\usepackage{gensymb}
\usepackage{amsmath}
\usepackage{graphics}

\begin{document}
\title{Robustness of superconductivity to competing magnetic phases in tetragonal FeS}
\date{\today}
\author{Franziska~K.~K.~\surname{Kirschner}}
\email{franziska.kirschner@physics.ox.ac.uk}
\affiliation{Department of Physics, University of Oxford, Clarendon Laboratory, Parks Road, Oxford, OX1 3PU, United Kingdom}
\author{Franz~\surname{Lang}}
\affiliation{Department of Physics, University of Oxford, Clarendon Laboratory, Parks Road, Oxford, OX1 3PU, United Kingdom}
\author{Craig~V.~\surname{Topping}}
\affiliation{Department of Physics, University of Oxford, Clarendon Laboratory, Parks Road, Oxford, OX1 3PU, United Kingdom}
\author{Peter~J.~\surname{Baker}}
\affiliation{ISIS Pulsed Neutron and Muon Facility, Rutherford-Appleton Laboratory, Chilton, Oxfordshire, OX11 0QX, United Kingdom}
\author{Francis~L.~\surname{Pratt}}
\affiliation{ISIS Pulsed Neutron and Muon Facility, Rutherford-Appleton Laboratory, Chilton, Oxfordshire, OX11 0QX, United Kingdom}
\author{Sophie~E.~\surname{Wright}}
\affiliation{Department of Chemistry, University of Oxford, Inorganic Chemistry Laboratory, South Parks Road, Oxford, OX1 3QR, United Kingdom}
\author{Daniel~N.~\surname{Woodruff}}
\affiliation{Department of Chemistry, University of Oxford, Inorganic Chemistry Laboratory, South Parks Road, Oxford, OX1 3QR, United Kingdom}
\author{Simon~J.~\surname{Clarke}}
\affiliation{Department of Chemistry, University of Oxford, Inorganic Chemistry Laboratory, South Parks Road, Oxford, OX1 3QR, United Kingdom}
\author{Stephen~J.~\surname{Blundell}}
\email{stephen.blundell@physics.ox.ac.uk}
\affiliation{Department of Physics, University of Oxford, Clarendon Laboratory, Parks Road, Oxford, OX1 3PU, United Kingdom}

\begin{abstract}
We have determined the superconducting and magnetic properties of a hydrothermally synthesized powder sample of tetragonal \ce{FeS} using muon spin rotation ($\mu$SR). The superconducting properties are entirely consistent with those of a recently published study, showing fully gapped behavior and giving a penetration depth of $\lambda_{ab} = 204(3)\,\rm{nm}$. However, our zero-field $\mu$SR data are rather different and indicate the presence of a small, non-superconducting magnetic phase within the sample. These results highlight that sample-to-sample variations in magnetism can arise in hydrothermally prepared phases, but interestingly the superconducting behavior is remarkably insensitive to these variations.
\end{abstract}

\maketitle

\section{Introduction}

Since the discovery of the first iron-based superconductors,\cite{Kami2006, Kami2008} this family of compounds has been a topic of intense interest. Initially, the focus was on iron arsenides for which the superconducting critical temperature $T_{\rm c}$ could reach 55\,K.\cite{Ren2008} Subsequently, it became possible to substantially enhance the superconducting properties of \ce{FeSe}\cite{Hsu2008} ($T_{\rm c} = 8\,{\rm K}$) using pressure,\cite{Medvedev2009} molecular intercalation,\cite{Burrard2013} and even via thin-film fabrication.\cite{Ge2015} Sulfides have received particular attention following the discovery of the first iron-sulfide superconductor, \ce{BaFe_2S_3}, reaching a $T_{\rm c}$ of 14\,K.\cite{Taka2015} The record for the highest $T_{\rm c}$ of any superconductor is currently held by a sulfide (203\,K for \ce{H_3S} at high pressure \cite{Droz2015}). Until recently, the possibility of studying superconductivity in the simplest sulfide analogue of iron selenide, FeS, had not been explored. This layered polymorph, mackinawite, is not trivial to synthesize, and had not previously been reported to be a superconductor. Recently, Lai \textit{et al.\ }reported a facile hydrothermal synthesis of this polymorph which also produced superconducting samples with a $T_{\rm c}$ of $\approx 5\,\rm{K}$.\cite{Lai2015} Pachmayr \textit{et al.\ }used single crystal x-ray diffraction measurements to show that such samples are stoichiometric \ce{FeS}.\cite{Pachmayr2016}

A variety of ground states have been predicted for \textit{t}-\ce{FeS}, including non-magnetic metallic,\cite{Devey2008, Borg2016, Subedi2008} and stripe antiferromagnetic order.\cite{Kwon2011} It is thought that the metallic state may arise due to delocalized iron $d$-electrons.\cite{Vaughan1971} It has been found that both the superconducting and magnetic properties of \textit{t}-\ce{FeS} are strongly dependent on fine details in the crystal structure.\cite{Kuhn2016} Density functional theory (DFT) calculations have shown that \textit{t}-\ce{FeS} is close to a spin-density wave (SDW) instability, and that the electronic structure and Fermi surface are very close to that of \ce{FeSe}.\cite{Subedi2008, Devey2008} The \ce{FeS_4} tetrahedra in \textit{t}-\ce{FeS} are closer to being perfectly regular\cite{Lai2015} than those in \ce{FeSe}\cite{Louca2010} (the S--Fe--S angle is $110.8(2) \degree$ compared to $104.02 \degree$ for Se--Fe--Se), and even though this normally favors superconductivity in arsenides,\cite{Lee2012} the value of $T_{\rm c}$ in \textit{t}-\ce{FeS} is lower than in \ce{FeSe}. It has been suggested that a low-moment magnetic phase with $T_{\rm N} \approx 20$\,K coexists with superconductivity in $t$-\ce{FeS}.\cite{PSI2016}  There has also been evidence that a magnetic anomaly exists below 15\,K, while commensurate antiferromagnetic order exists below 116\,K.\cite{Kuhn2016}

In this paper, we perform muon spin rotation experiments on a powder sample of $t$-\ce{FeS}, in order to determine its magnetic and superconducting properties. Our results show that while the superconducting properties of our samples match those in previous studies, the magnetic behaviour is rather different, highlighting both the robustness of the superconducting state to magnetic disorder and the sensitivity of the magnetism to details of sample preparation.

\section{Sample preparation}

A sample of \textit{t}-\ce{FeS} was synthesized via a slightly modified literature procedure.\cite{Lai2015} Elemental \ce{Fe} powder (ALFA, 99.998\%) and anhydrous \ce{Na_2S} (synthesized by the reaction of elemental sodium and sulfur in the correct ratio in liquid ammonia at $-$50\,\degree C) were weighed out in a 1:1 molar ratio and sealed inside a Teflon lined steel autoclave after being solvated with deionised water (10\,ml). The reaction was heated at 120\,\degree C for 6 days before being furnace cooled to room temperature. The material was washed with deionised water and dried under vacuum. X-ray powder diffraction measurements showed the presence of elemental Fe within the sample. As such the isolated sample was reacted with a further 0.5 molar equivalents of anhydrous \ce{Na_2S} in a Teflon lined steel autoclave for 3 days at 130\,\degree C before being furnace cooled to room temperature. The dark grey powder was then washed with deionised water and dried under vacuum. Magnetometry measurements (using a Quantum Design SQUID magnetometer) confirmed that the sample was superconducting with a $T_{\rm c} \approx 4\,\rm{K}$ (see Fig.~\ref{TFfit}(a)).

\section{$\mu$SR experiments}

$\mu$SR experiments\cite{Blundell1999, Yaouanc2011} were performed using a \ce{He^3} cryostat inserted in the MuSR spectrometer at the ISIS pulsed muon facility, Rutherford Appleton Laboratory, UK.\cite{King2013}  Initially, transverse-field (TF) measurements were performed, in which a field is applied perpendicular to the initial direction of muon polarization to determine the nature of the superconductivity and the critical temperature. Next, zero-field (ZF) measurements were taken to determine which, if any, competing magnetic phases were present in the sample. For both sets of measurements, the sample was mounted on a haematite backing plate in order to remove the background signal which arises from the sample holder, causing a reduction in the oscillating (TF) and relaxing (ZF) amplitude in our data. All of the data were analyzed using WiMDA.\cite{WiMDA}

\subsection{TF measurements}

\begin{figure}[t]
\centering
\includegraphics[width=.5\textwidth]{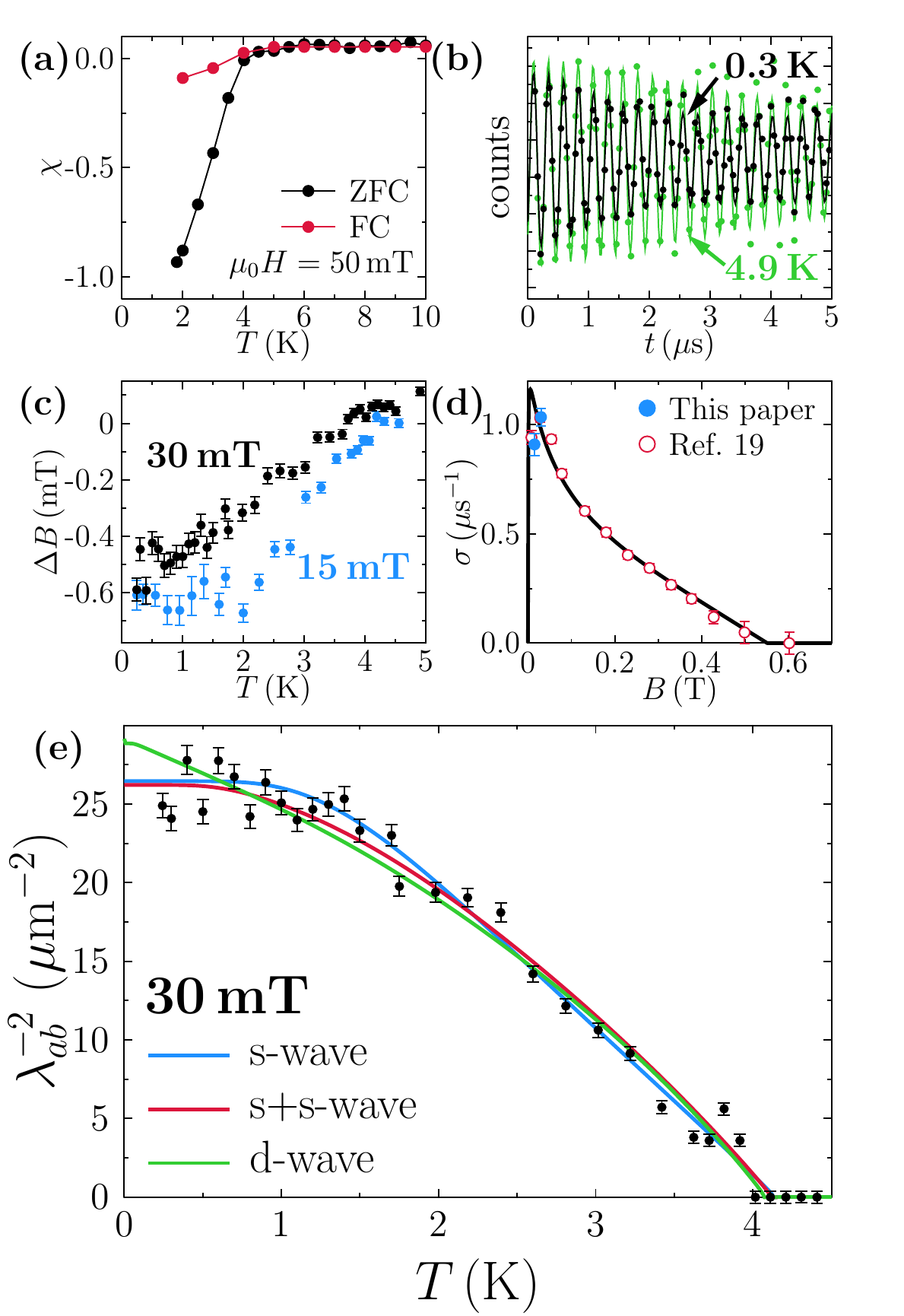}
\caption{(a) The field cooled (FC) and zero-field cooled (ZFC) magnetic susceptibility, $\chi=M/H$ where $M$ is the sample's magnetization, for \textit{t}-\ce{FeS} (given in dimensionless units) in a magnetic field of $\mu_0H=50\,\rm {mT}$. (b) Sample TF-$\mu$SR spectra above and below $T_{\rm c}$ at an applied field of 30\,mT, where the raw counts from one set of detectors has been plotted. (c) The shift in magnetic field $\Delta B = B_{\rm SC} - B_{\rm app}$ due to the superconducting vortex lattice in applied fields of 15\,mT and 30\,mT. (d) Zero-temperature relaxation for our measurements, and those from Ref.~\onlinecite{PSI2016}, fitted to Eq.~\ref{siglambda} with a low-field correction factor (black line). (e) The inverse square perpendicular penetration depth $\lambda_{ab}^{-2}$ at an applied field of 30\,mT, calculated from TF-$\mu$SR spectra fitted with Eq.~\ref{TFeqn}. Superfluid density functions for $s$-wave, $s+s$-wave and $d$-wave superconductivity have been fitted as in Eq.~\ref{BCSfit}.}
\label{TFfit}
\end{figure}

TF measurements above and below $T_{\rm c}$ were performed in magnetic fields $B_{\rm app}$ of 30\,mT and 15\,mT between temperatures $T$ of 0.24\,K and 4.9\,K, and sample spectra are given in Fig.~\ref{TFfit}(b). There is a clear increase in relaxation below $T_{\rm c}$, caused by the onset of the superconducting vortex state, which produces an inhomogenous magnetic field distribution inside the sample \cite{Brandt1988}. The data were fitted with the two-component function
\begin{multline} \label{TFeqn}
A(t) = A_{B} \cos\left(\gamma_{\mu} B_{\rm app}t + \phi \right) \exp\left[ - \lambda_{\rm TF} t\right] \\
+ A_{\rm SC} \cos\left(\gamma_{\mu} B_{\rm SC} t + \phi \right) \exp\left[ - \left(\sigma t\right)^2/2\right],
\end{multline}
where $\gamma_{\mu} = 2 \pi \times 135.5\,{\rm MHzT}^{-1}$ is the gyromagnetic ratio of the muon and $\phi$ is related to the detector geometry (the data were divided among eight groups of detectors with $\phi$ fitted for each group). The first term corresponds to muons that do not experience any superconductivity, and precess in the external field. There exists a weak exponential relaxation from magnetism in the sample (see Section~\ref{ZF}) and also a small contribution from muons implanted in the cryostat. The second term corresponds to the superconducting fraction of the sample, and the Gaussian broadening $\sigma(T) = \sqrt{\sigma_{\rm SC}^2(T) + \sigma_{\rm nucl}^2}$ contains contributions from the vortex lattice (which is temperature dependent) and the nuclear moments; we find $\sigma_{\rm nucl} = 0.368(3)\,\mu\rm{s}^{-1}$ and hence can deduce $\sigma_{\rm SC}$ from fitted values of $\sigma$.

Figure~\ref{TFfit}(c) shows that there is a clear small negative shift in the average field due to the vortex lattice, $\Delta B~=~B_{\rm SC}-B_{\rm app}$ caused by the distribution of fields of the vortex lattice (whose average field is slightly lower than the applied field), which decreases as $T_{\rm c}$ is approached and vortices disappear.

As the sample is anisotropic and polycrystalline, it can be assumed that the effective penetration depth $\lambda_{\rm eff}$ is dominated by the in-plane penetration depth $\lambda_{ab}$ since $\lambda_{ab} \ll \lambda_{c}$, and so $\lambda_{\rm eff} = 3^{1/4} \lambda_{ab}$.\cite{Fesenko1991} Assuming \textit{t}-\ce{FeS} is type II superconductor with an isotropic hexagonal Abrikosov vortex lattice in the $a$--$b$ plane that can described by Ginzburg-Landau theory, the relaxation due to the vortex lattice can be related to the penetration depth by the approximation\cite{Brandt2003}

\begin{equation} \label{siglambda}
\sigma_{\rm SC}(T) = 0.0993 \frac{\gamma_{\mu} \phi_0}{2\pi}  \left(1-b\right) \left(1+1.21 \left( 1-\sqrt{b} \right)^3 \right)  \lambda_{ab}^{-2}(T),
\end{equation}
where $b = B_{\rm app}/B_{\rm c2}$ is the reduced field, and $\phi_0 = 2.069 \times 10^{-15}\,{\rm Wb}$ is the magnetic flux quantum. $\sigma_{\rm SC}$ is given in units of $\mu$s$^{-1}$ and $\lambda_{ab}^{-2}$ in units of $\mu$m$^{-2}$. This expansion holds to within 5\% for $\kappa \geq 5$ and $0.25/\kappa^{1.3} < b < 1$, where the Ginzburg-Landau parameter of \textit{t}-\ce{FeS} $\kappa \approx 9$.\cite{PSI2016} 

The behaviour of $\sigma_{\rm SC}$ can be extended to lower fields using an additional correction to Eq.~(\ref{siglambda}) that takes account of the failure of the approximation $G\lambda \gg 1$ in the London formula $(\delta B)^2 = \sum_{\textbf{G} \neq 0} B_{\rm app}^2/(1+G^2\lambda_{\rm eff}^2)^2$, where $\{ \textbf{G} \}$ is the set of reciprocal lattice vectors.\cite{Brandt2003} The width of the field distribution is related to the relaxation by $\delta B = \sigma_{\rm SC} / \gamma_{\mu}$. Figure~\ref{TFfit}(d) shows the product of this correction factor (which $\approx 1$ for $B_{\rm app} \gtrsim 0.25B_{\rm c2}/\kappa^{1.3}$) and Eq.~(\ref{siglambda}). Our data only covers the low-field region, but is entirely consistent with that obtained in Ref.~\onlinecite{PSI2016}. Fitting both datasets with this correction to Eq.~\ref{siglambda} yields a value of the upper critical field $B_{\rm c2} = 0.55(3)\,{\rm T}$ and penetration depth $\lambda_{ab}(0) = 204(3)\,{\rm nm}$.

The parameter $\Delta B$ plotted in Fig.~\ref{TFfit}(c) is governed by the penetration depth: in \textit{t}-\ce{FeS} this is relatively large, resulting in the small $\Delta B$. The low-temperature measured $\Delta B$ values are close to the theoretical predictions: $\Delta B_{30\,{\rm mT}} \approx -0.93\,\rm{mT}$ and  $\Delta B_{15\,{\rm mT}} \approx -0.97\,\rm{mT}$ as calculated using the approximation\cite{Brandt2003}
\begin{equation}
\Delta B \approx - 0.146 B_{\rm c2} \frac{1-b}{\kappa^2 - 0.069}.
\end{equation}
The experimental values are slightly lower in magnitude than the theory, and this can be explained by the Gaussian approximation made when fitting the vortex lattice field distribution in Eq.~(\ref{TFeqn}): in reality, the field distribution is not symmetric and is skewed towards lower fields, resulting in a larger shift.\cite{Brandt1988}

Figure~\ref{TFfit}(e) shows $\lambda_{ab}^{-2}$ as a function of temperature at 30\,mT (very similar results were obtained for 15\,mT). The data have been fitted with BCS single- and two-gap $s$-wave models, and a single-gap $d$-wave model. The BCS model of the normalized superfluid density of a superconductor is given by \cite{Chand1993}
\begin{align} \label{BCSfit}
\tilde{n}_{\rm s}(0) & = \frac{\lambda^{-2}(T)}{\lambda^{-2}(0)} \notag\\
& = 1 + \frac{1}{\pi} \int_0^{2\pi} \int_{\Delta(\phi,T)}^{\infty} \frac{\partial f}{\partial E} \frac{E\, {\rm d}E {\rm d}\phi}{\sqrt{E^2 - \Delta^2(\phi,T)}} ,
\end{align}
where $\Delta(\phi,T)$ is the superconducting gap function and $f = \left(1+\exp\left(E/k_{\rm B}T\right)\right)^{-1}$ is the Fermi function. The gap function can be approximated by $\Delta(\phi,T) = \Delta(\phi)\tanh\left[1.82\left(1.018\left(T_{\rm c}/T - 1\right)\right)^{0.51} \right]$ where the angular function $\Delta(\phi) = \Delta_0$ in the $s$-wave model and $\Delta(\phi) = \Delta_0\cos(2\phi)$ in the $d$-wave model,\cite{Carrington2003} where $\Delta_0$ is a constant. The two-gap $s$-wave model is a simple superposition of two single-gap $s$-waves weighted by a factor $w$: $\tilde{n}_{\rm s}(T) = w\tilde{n}_{\rm s}^{(1)}(T) + (1 - w)\tilde{n}_{\rm s}^{(2)}(T)$.

The fitting favors fully-gapped behavior but cannot easily distinguish whether a single- or two-gap $s$-wave model is more appropriate. The $s$-wave model with $\Delta_0 =  0.52(1)\,\rm{meV}$ gives the lowest overall $\chi^2$, and so this is taken as the model of best fit, although there is little difference between this and the  $s+s$-wave model. This conclusion contrasts with some other studies of \textit{t}-\ce{FeS}, which have found evidence for nodes in the gap function.\cite{Ying2016, Xing2016} However, a previous $\mu$SR study has also reported fully-gapped behaviour, though the best fit was for two $s$-wave gaps.\cite{PSI2016}

Our extracted critical temperature $T_{\rm c} = 4.13(3)\,\rm{K}$ at 30\,mT is in close agreement with previous studies on \textit{t}-\ce{FeS}.\cite{Lai2015, Kuhn2016, PSI2016, Xing2016, Zhang2016}

\subsection{ZF measurements} \label{ZF}

\begin{figure}[t]
\centering
\includegraphics[width=.5\textwidth]{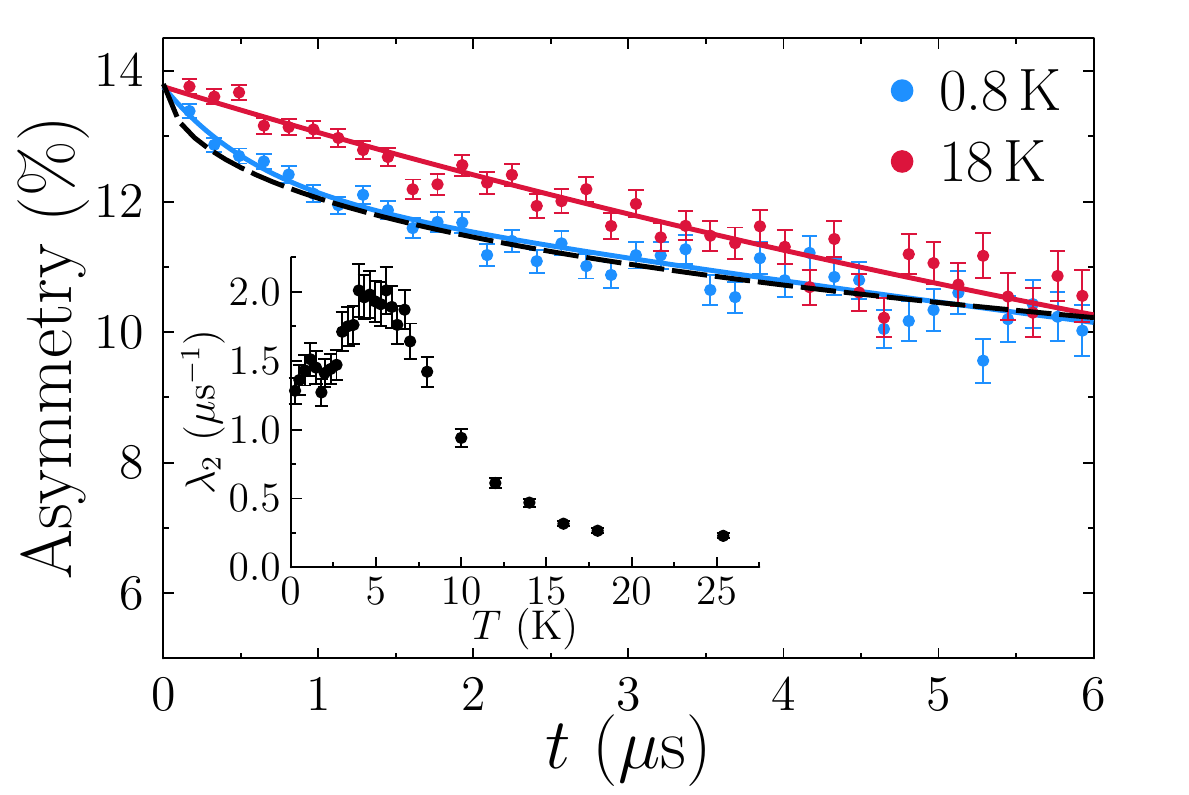}
\caption{Sample ZF-$\mu$SR spectra for \textit{t}-\ce{FeS} below and above $T_{\rm c}$. Two-component exponential fits, given by Eq.~\ref{Lor}, are also plotted. The black dashed line shows a fit from a simulated dilute, dynamic spin system. The inset shows the temperature dependence of the relaxation rate of the fast-relaxing phase in Eq.~\ref{Lor}.}
\label{ZFfig}
\end{figure}

Despite the remarkable similarity in the superconducting properties between our sample and that reported in Ref.~\onlinecite{PSI2016}, the magnetic properties have been found to be markedly different. Sample ZF-$\mu$SR spectra above and below $T_{\rm c}$ are given in Fig.~\ref{ZFfig}. We observe a slight increase of relaxation as $T$ approaches $T_{\rm c}$, but this is not as drastic a change as reported in Ref.~\onlinecite{PSI2016}. [Note that in Fig.~\ref{ZFfig} the initial ($t=0$) asymmetry is only $\approx 14\%$, lower than the maximum expected for data on this spectrometer, largely due to muons being absorbed by the haematite.] No oscillations were seen in the forward-backward asymmetry spectra, and there were also no discontinuous jumps in either the initial or baseline asymmetry. This, combined with the absence of a Kubo-Toyabe relaxation (ruling out effects from nuclear moments), hints that dynamic moments with no long-range order exist in the sample. This can be modelled by a two-component exponential relaxation
\begin{equation} \label{Lor}
A(t)=A_{1}\exp\left(-\lambda_{1} t\right) + A_2\exp\left(-\lambda_2 t\right),
\end{equation}
which takes into account a slowly-relaxing background with amplitude $A_{1}$ and relaxation rate $\lambda_{1}$, and a further signal with amplitude $A_2$ and a faster relaxation rate $\lambda_2$. The total observed amplitude $A_0$ with baseline $A_{\rm base}$ comprises of these two components, such that $A_0-A_{\rm base} = A_{1} + A_2$. Exponential relaxation corresponds to either dynamic moments with a single correlation time within the resolution of the spectrometer and an unknown field distribution,\cite{Khasanov2008} or a dilute distribution of static moments.\cite{Walstedt1974} The slower background relaxation (with constant $\lambda_{1} \approx 0.04\,\mu\rm{s}^{-1}$) could arise due to the intrinsic magnetic moments of the iron in \textit{t}-\ce{FeS} (which is in contrast to behaviour observed in \ce{FeSe} \cite{Khasanov20082}), whereas the faster relaxation may be due to a magnetic impurity phase. The inset in Fig.~\ref{ZFfig} shows the temperature dependence of the faster relaxation rate $\lambda_2$, measured between 0.24\,K and 25.3\,K. As temperature decreases, the relaxation rate increases until $T \approx 5\,{\rm K}$ and drops slightly thereafter. This is characteristic for a magnetic phase that coexists and competes with a superconducting phase. As the peak coincides closely with the superconducting transition in \textit{t}-\ce{FeS}, it is likely that the onset of magnetism is coupled to the superconducting order parameter. It should be noted that at all measured temperatures $\lambda_2$ is large, but the relative amplitude $A_2/(A_1 + A_2) \approx 15\%$ is small, indicating a very low density of moments with a large distribution of stray fields.

We have carried out simulations to explore the effect of a system of dilute, dynamic spins which could arise from localized magnetic impurities within the \ce{FeS} layers.  Our results are consistent with experimental observations if the concentration of such impurities is $\sim 1\%$ and the fluctuation rate of these spins is $\approx 0.1$--$0.2\,\rm{GHz}$ (this is shown in the black line plotted in Fig.~\ref{ZFfig}), although it is possible to achieve good agreement with higher concentrations of impurity spins with correspondingly higher fluctuation rates.  However, our x-ray diffraction data rule out any magnetic impurity phase at a higher concentration than the per cent level, and lower concentrations have been excluded by the simulations (as they do not agree with the data). Simulations of static spin distributions also did not fit the data and so we conclude that the fluctuations of these dilute spins are important (the situation is reminiscent of effects observed in spin glasses\cite{Uemura1985}). The moments were found to be on the order of $\approx 1\mu_{\rm B}$, which suggests the magnetic phase could arise due to atomic iron or iron-based impurities in the \ce{FeS} layers, similar to those found in \ce{FeSe}.\cite{Hsu2008, Hu2011}  We think this more likely than the low-moment ($10^{-2}$--$10^{-3}\,\mu_{\rm B}$) uniform magnetic phase of uncertain origin postulated by Ref.~\onlinecite{PSI2016}.

The ZF results obtained in this experiment are significantly different to those in previous $\mu$SR work on \textit{t}-\ce{FeS}. In Ref.~\onlinecite{PSI2016}, a magnetic transition was observed at $T \approx 20\,{\rm K}$, resulting in a sharp increase in the relaxation rate, peaking at $\lambda \approx 0.5\,\mu{\rm s}^{-1}$ with the relative fraction of the magnetic phase $\approx 89 \%$. This is markedly different to the behaviour observed in our sample, and suggests that the magnetic properties of \textit{t}-\ce{FeS} are strongly sample-dependent, a conclusion which is supported by Ref.~\onlinecite{Kuhn2016}.

The similarity of the superconducting properties demonstrates that the superconducting ground state is remarkably insensitive to these differences in magnetism. Moreover, this supports the conclusion that the gap is isotropic because a ground state with nodes might be expected to be far more susceptible to the presence of impurities, which may be the source of these sample-dependent differences.

\section{Conclusions}

In summary, we have carried out TF- and ZF-$\mu$SR experiments on a sample of \textit{t}-\ce{FeS} to determine its superconducting and magnetic properties. It was found that the superconductivity in \textit{t}-\ce{FeS} agrees remarkably well with both previous experiments and theoretical predictions, and shows the robustness of the superconducting phase. The magnetic properties were found to be significantly different to those measured in other samples, which demonstrates the high sensitivity of the magnetic phase to small alterations in the synthesis process, and moreover it highlights the insensitivity of the superconducting phase to these changes. It is anticipated that the $T_{\rm c}$ of \textit{t}-\ce{FeS} could be enhanced with similar chemical modifications to \ce{FeSe}. Our results support the notion that if $t$-\ce{FeS} can be chemically modified using techniques similar to those that have led to an enhancement of $T_{\rm c}$ in \ce{FeSe} intercalates, then the superconductivity of \ce{FeS} layers may prove to be remarkably resilient.

\section{Acknowledgements}

We thank ESPRC and the Leverhulme Trust (RPG-2014-221) for funding support and P.~Biswas for experimental support. F.\ K.\ K.\ K.\ thanks Lincoln College Oxford for a doctoral studentship. Part of this work was performed at the Science and Technology Facilities Council (STFC) ISIS facility, Rutherford Appleton Laboratory.

\end{document}